\documentclass[pre,aps,showpacs,twocolumn,floatfix]{revtex4}
\usepackage{graphicx}
\usepackage{epsfig,graphics}
\usepackage{amsmath, amsthm, amssymb}
\usepackage{color}

\newcommand{\be}{\begin{equation}}
\newcommand{\ee}{\end{equation}}
\newcommand{\bea}{\begin{eqnarray}}
\newcommand{\eea}{\end{eqnarray}}

\newcommand{\pt}{\tilde{p}}
\newcommand{\li}{\operatorname{Li}}

\begin{document}
\title{Asymptotic Behavior of Inflated Lattice Polygons}
\author{Mithun K. Mitra}
\email{mithun@imsc.res.in} 
\affiliation{The Institute of Mathematical Sciences, 
C.I.T. Campus, Taramani, Chennai 600113, India}
\author{Gautam I. Menon}
\email{menon@imsc.res.in} 
\affiliation{The Institute of Mathematical Sciences, 
C.I.T. Campus, Taramani, Chennai 600113, India}
\author{R. Rajesh}
\email{rrajesh@imsc.res.in}
\affiliation{The Institute of Mathematical Sciences, 
C.I.T. Campus, Taramani, Chennai 600113, India}
\date{\today}

\begin{abstract}

We study the inflated phase of two dimensional
lattice polygons with fixed perimeter $N$ and
variable area, associating a weight $\exp[pA - Jb
]$ to a polygon with area $A$ and $b$ bends. For
convex and column-convex polygons, we show
that $\langle A \rangle/A_{max} = 1 - K(J)/\pt^2
+ \mathcal{O}(\rho^{-\pt})$, 
where $\pt=pN \gg 1$, and $\rho<1$.
The constant $K(J)$ is found to be the same
for both types of polygons. We argue that
self-avoiding polygons should exhibit the same
asymptotic behavior. For self-avoiding polygons,
our predictions are in good agreement with 
exact enumeration data for $J=0$ and Monte Carlo
simulations for $J \neq 0$. We also study polygons
where self-intersections are allowed, verifying numerically
that the asymptotic behavior described above continues
to hold.

\end{abstract}


\maketitle

\section{\label{intro} Introduction}

The study of lattice polygons weighted by area
and perimeter is a central problem in lattice
statistics and combinatorics. Lattice polygons have
been used to model vesicles \cite{leibler87,fisher91},
cell membranes \cite{satyanarayana04}, emulsions
\cite{faassen98}, polymers \cite{privman} and
percolation clusters \cite{rajesh05}. In several cases, exact
generating functions for classes of such polygons have been
obtained. A survey of different kinds
of lattice polygons and a review of related results
can be found in Refs.~\cite{bousquet96,rensburgbook}.

In this paper, we study the asymptotic behaviour
of the area enclosed by inflated polygons of fixed
perimeter. We calculate the area
for two special cases of lattice
polygons - convex and column-convex lattice polygons. We then
conjecture the appropriate form for the area of
self-avoiding polygons in the inflated phase.

We first summarize known results for the
problem of pressurized polygons, based on 
the generating function
\be
G(\mu,p)  = \sum_{A,N} C_N(A) e^{p A} \mu^N,
\ee
where $C_N(A)$ is the number of self-avoiding polygons
of perimeter $N$ and area $A$, weighted by a chemical
potential $\mu$. Here $p$ is the 
pressure which couples to the area $A$.
Exact solutions exist for $G(\mu,p)$
when $C_N(A)$ is restricted to convex polygons
\cite{lin91,bousquet92_1,bousquet92_2} or to
column-convex polygons \cite{brak90_1}. However,
a general solution for self-avoiding polygons is unavailable.
Exact enumeration results for self-avoiding polygons exist for all $N$ up to
$N=90$ and for all $A$ for these values of $N$ \cite{jensen03}.  
A transition
at $p=0$ separates a branched polymer phase when $p<0$
(for $\mu$ sufficiently small) from an inflated phase
when $p>0$.  At $p=0$, the problem reduces to that of
the enumeration of self-avoiding polygons. The scaling
function describing the scaling behavior (for $p<0$)
near the tricritical point $p=0$ and $\mu=\kappa^{-1}$,
where $\kappa$ is the growth constant for self-avoiding
polygons, is also known exactly \cite{richard01,cardy01,richard02}.

Less is known about the inflated phase obtained for positive pressures 
$p > 0$.  However,
this phase is of physical interest in the case of
two-dimensional vesicles, or equivalently pressurized ring polymers
\cite{leibler87,rudnick91,gaspari93,haleva06,mitra07}.
In the calculations described in this paper, 
we consider the partition function
\be
\mathcal{Z}_N(p,J) = \sum_{A,b} C_N(A,b) e^{p A - Jb}, \quad p>0,
\ee
where $C_N(A,b)$ is the number of self-avoiding polygons of 
area $A$ with $b$ bends.  A bending energy 
cost $J$ per bend is introduced to incorporate semi-flexibility.

Some rigorous results exist for $Z_N(p,0)$ when $p>0$. 
Ref.~\cite{prellberg99} proved that
\be
\mathcal{Z}_N(p,0)= A(p) e^{p N^2/16} (1+\mathcal{O}(\rho^N)) ~~\mathrm{as}~~
N\rightarrow \infty ,
\label{prellbergresult}
\ee
for some $0<\rho<1$, with $A(p)$ some function of $p$. 
This result holds in the limit where $N\rightarrow \infty$
keeping $p$ fixed.
However, as we argue in Sec.~\ref{sec2}, the relevant scaling limit 
in the inflated regime is $p \rightarrow 0$, 
$N \rightarrow \infty$ keeping $\pt=pN$ finite.

The central result of this paper is 
then the following: In this limit, 
we show that for both convex and column-convex polygons, the
area is given
by,
\be
\langle A \rangle= \frac{N^2}{16} \left[1 -\frac{32 \pi^2}{3 \pt^2} + 
\frac{64}{\pt^2} 
\mathrm{Li}_2\left(1-\alpha \right) \right] + \mathcal{O}(e^{-\pt/8}) ,
\ee
where $\li_2$ is the dilogarithm function
\be
\li_2 (x) = \sum_{m=1}^{\infty} \frac{x^m}{m^2} ,
\ee
and, $\alpha = e^{-2J}$. 
We argue that this result should also extend to the self-avoiding
case and test this conjecture numerically.

The paper is organized as follows. In Sec.~\ref{sec2},
we present a justification of the scaling limit we consider
using a simply Flory-type argument.
Sections.~\ref{sec3} and \ref{sec4} 
contain the calculation of the area for convex and column-convex polygons
respectively. Section~\ref{sec5} contains the numerical analysis of
self-avoiding polygons and self
intersecting polygons. A brief summary of our results 
and conclusions is presented in Sec.~\ref{sec6}.

\section{\label{sec2} Scaling in the inflated regime}

The inflated regime of self-intersecting pressurized polygons 
has been well studied in the continuum 
\cite{rudnick91,gaspari93,haleva06,mitra07}. In this case
the appropriate scaling variable is obtained
by scaling the thermodynamic pressure with the system size,
taking $p \rightarrow 0$, 
$N \rightarrow \infty$ keeping $\pt=pN$ finite. A typical
configuration in the inflated phase has  no self-intersections.
Thus, we expect that the above scaling should also hold
for self-avoiding polygons. 

To motivate our choice of the scaling variable, 
we shall follow the Flory-type scaling analysis developed in 
Ref.~\cite{maggs90}. 
The free energy in the inflated phase consists of 
two terms. These  describe the contribution from the
pressure differential and the stretching free energy 
of the closed self-avoiding walk. The pressure contribution 
is
\be
F_{pressure} = -p A \approx -p R^2 ,
\ee
where it is assumed that the area scales as the square 
of the radius of the ring. The stretching free energy is
\be
F_{stretching} \approx \frac{R^4}{N^3} ,
\ee
in two-dimensions \cite{maggs90,fisher66}. The total free energy
is then given by the sum of the two contributions,
\be
F = F_{pressure} + F_{stretching} . 
\ee
Thus, in the inflated phase, the radius of the ring
scales as
\be
\langle R^2 \rangle \sim N^3 p .
\ee
Now, in the inflated phase, $\langle R^2 \rangle$ should scale 
as the square of the number of monomers. This implies 
the following scaling
\be
\langle R^2 \rangle \sim N^2 \pt ,
\ee
with $\pt = p N$. 

\begin {figure}
\includegraphics[width=\columnwidth]{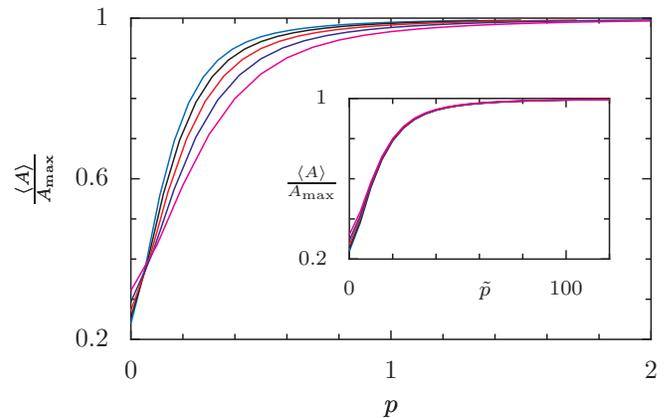}
\caption{\label{scaling} The variation of area with pressure $p$ for
self avoiding polygons on a square lattice.
Inset: When plotted as a function of the scaling variable $\pt=pN$,
the area curves for different values of $N$ collapse onto each other.
The system sizes used are $N=50,60,70,80,90$.
The data is generated from exact enumerations of the polygons on the square lattice
\cite{jensensap}.}
\end {figure}

In Fig.~\ref{scaling}, we show the variation of $\langle A \rangle/A_{max}$ 
with pressure $p$, where $A_{max} = N^2/16$ is the maximum possible area. The
data points collapse onto one curve when $p$ is scaled as $\pt=pN$.
The data is
obtained from exact enumerations of self-avoiding polygons on the 
square lattice \cite{jensensap}. 

\section{\label{sec3} Convex Polygons}

\begin {figure}
\includegraphics[width=\columnwidth]{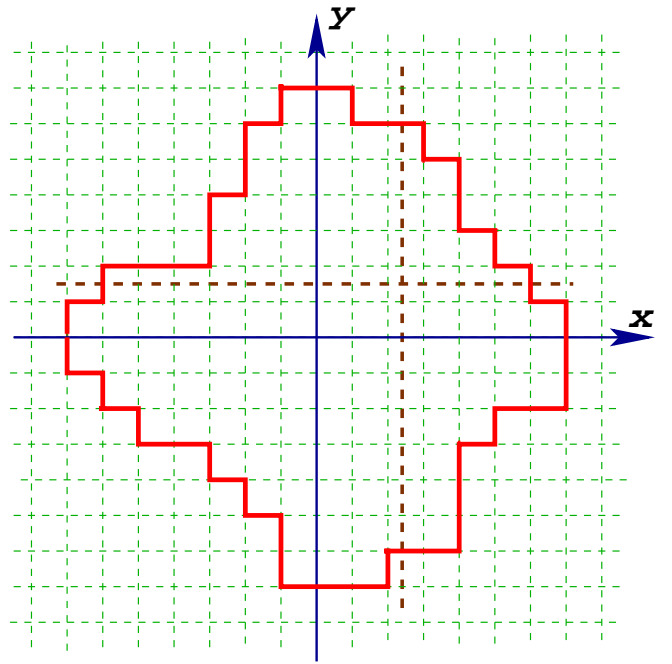}
\caption{\label{convexfig} A schematic diagram of a convex polygon. 
Any vertical or horizontal line (thick dashed lines) 
intersects the convex polygon at either 
$0$ or $2$ points.  }
\end {figure}

In this section we calculate the equilibrium shape and area
of a convex polygon when $\pt > 0$.
Convex polygons are those polygons which have exactly $0$ or $2$
intersections with any vertical or horizontal line drawn through the
midpoints of the edges of the lattice (see
Fig.~\ref{convexfig}). We calculate the area by determining the shape of
the convex polygon that minimizes the free energy at fixed 
perimeter, generalizing the calculation presented in Ref.~\cite{rajesh05}.

The perimeter $N$ of a convex polygon is the same as that of its bounding box,
which, in general, is a rectangle. The equilibrium shape should however
be invariant about rotations by angle $\pi/2$. The bounding box of the
equilibrium shape is thus a square of side $N/4$. We can now calculate
the shape in the first quadrant,  obtaining the shapes in other quadrants by
symmetry.

Consider a coarse grained shape $y(x)$ in the first quadrant  
with endpoints at $(0,N/8)$ and
$(N/8,0)$. The free energy functional for this curve $y(x)$ can be written as
\be
\mathcal{L}[y(x)] = \int_0^{N/8} dx \, \sigma(y')\sqrt{1+y'^2} - \frac{\pt}{N}
\int_0^{N/8} dx~ y ,
\label{lagconvex}
\ee
where $\sigma(y')$ is the free energy per unit length associated with a
slope $y'$ and $\pt$ is the scaled pressure. The shape is then obtained 
obtained from Eq.~(\ref{lagconvex}) through the Euler Lagrange equation (Wulff
construction)
\cite{rottmannwortis84},
\be
\frac{d}{d x} \frac{d}{d y'} \left[ \sigma(y') \sqrt{1+y'^2} \right] = 
-\frac{\pt}{N}.
\label{eq:euler}
\ee
The free energy can be calculated using  a simple combinatorial argument.
Consider all possible paths starting from $(0,y)$ and ending at 
$(x,0)$ with only rightward and downward steps. The weight of a path is
$\exp(-J b)$, where $b$ is the number of bends.
When $x,y \gg 1$, the weighted sum of these paths will be equal to
$\exp[- \sqrt{x^2+y^2} \sigma (y')]$, where $y' = -y/x$. 

Let $C(x,y)$ be the sum of weighted walks constructed as above.
Such walks may be enumerated by splitting the path into
sequences of rightward and downward steps and associating the
bending energy term to a sequence of downward ($y$) steps
begun and terminated by a step to the right.
Then 
\be
C(x,y) = \sum_{y_1,y_2,\ldots,y_X} \prod_i W_i ~\delta\left[\sum y_i - y\right] ,
\ee
where $W_i$ is the weight associated with the $i^{th}$ step,
which is given by 
\be
W_i = [ \alpha (1-\delta_{y_i,0}) + \delta_{y_i,0} ] ,
\ee
and $\alpha = e^{-2 J}$.
The delta function enforces the constraint that the steps taken in the
$y$-direction must total $y$. The summation is over all possible numbers
of steps taken in the $y$ direction at steps $1, 2 \ldots x$.

Performing a discrete Laplace transform, we obtain
\bea
\sum_y C(x,y) \omega^y &\approx& \left[ 1 + \alpha \omega + \alpha \omega^2 +
\cdots \right]^x, \nonumber \\
&=& \left[\frac{1 - (1-\alpha) \omega}{1-\omega} \right]^x.
\eea
For large $x,y$, the inverse Laplace transform can be calculated
using the saddle point approximation. This gives
\be
\sigma(y') = \frac{-f(\omega^*)}{\sqrt{1+y'^2}},
\label{eq:surfacetension}
\ee
where 
\be
f(\omega) = y' \ln(\omega) + \ln[1-(1-\alpha) \omega] - \ln (1-\omega),
\ee
and $\omega^*$ satisfies
\be
\left. \frac{d f}{d \omega} \right|_{\omega^*} = 0.
\ee

\begin {figure}
\includegraphics[width=\columnwidth]{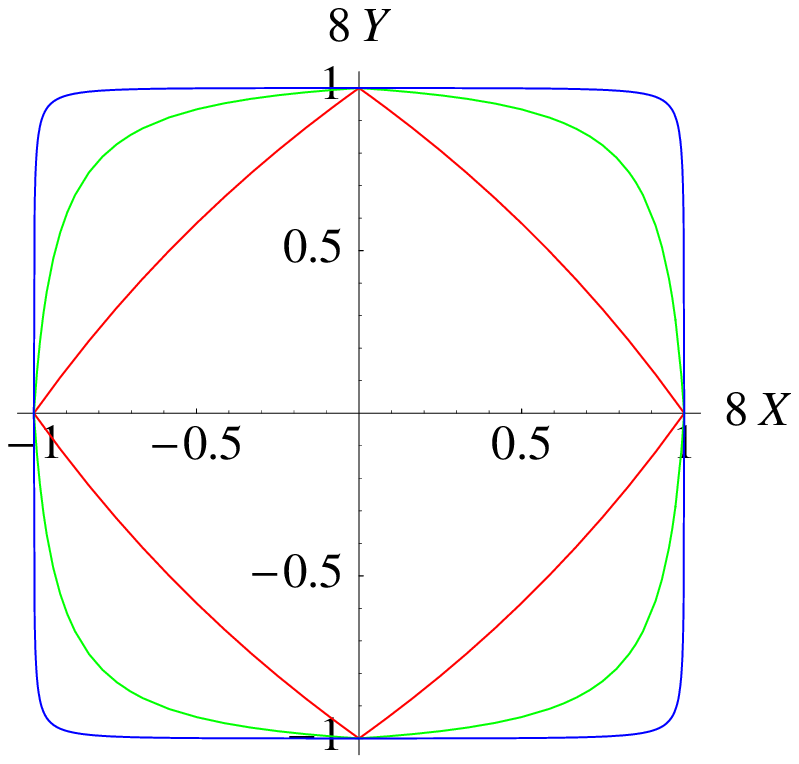}
\caption{\label{convexshape} Shape of the convex polygon as obtained from
Eq.~(\ref{convexshapeeq}) for polygons with $J=1$. The different 
shapes correspond to pressure values $\pt=1.0,10.0,50.0$, with the outer
shape corresponding to largest pressure.}
\end {figure}

The equilibrium shape can now be obtained from Eqs.~(\ref{eq:euler})  and
(\ref{eq:surfacetension}). The shape satisfies the equation
\be
c e^{\pt X} + c c_1 e^{\pt (X+Y)} (1-\alpha) - c_1 e^{\pt Y} = 1 , 
\ee
where $X$ and $Y$ are scaled coordinates defined as $X=x/N$ and $Y=y/N$.
The constants of integration are fixed by imposing the requirement that the
shape should be symmetric under the interchange of $X$ and $Y$, and 
the boundary 
condition that $y(x=N/8)=0$. This gives,
\be
c = \! -c_1 \! = 
\frac{(1+e^{-\pt/8}) - \sqrt{(1+e^{-\pt/8})^2 - 4 e^{-\pt/8} (1-\alpha)}}
{2 (1-\alpha)} ,
\label{eq:cc1}
\ee
and the equilibrium shape can be written as
\be
c e^{\pt X} - c^2 e^{\pt (X+Y)} (1-\alpha) + c e^{\pt Y} = 1 .
\label{convexshapeeq}
\ee
The shapes for different values of the scaled pressure are shown in 
Fig.~\ref{convexshape} for
a convex polygon with $J=1$.

The area of the convex polygon is  obtained from the
equilibrium shape as 
\be
\langle A \rangle = 4 N^2 \int_0^{1/8} Y dX , 
\ee
where the factor of $4$ corresponds to  the 
four quadrants. Doing the integration, 
we obtain
\bea
\langle A \rangle &=& \frac{N^2}{16} \left[ - \frac{8 \ln(c)}{\pt} + 
\frac{64}{\pt^2}
\left(  \mathrm{Li}_2[c] 
-\mathrm{Li}_2[c~e^{\pt /8}] \right. \right. \nonumber \\
&+& \left. \left. \mathrm{Li}_2[c(1-\alpha) e^{\pt/8} ] 
-\mathrm{Li}_2[c(1-\alpha) ]
\right)
\right] ,
\label{eq:convexarea}
\eea
where $\li_2$  is the dilogarithm function.

The asymptotic behavior for large $\pt$ may be calculated from 
Eqs.~(\ref{eq:cc1}) and (\ref{eq:convexarea}).
When $\pt \gg 1$, the constant $c$ can be written as, 
\be
c = e^{-\pt/8} + \mathcal{O}(e^{-\pt/4}) . 
\ee
Substituting into Eq.~(\ref{eq:convexarea}), we obtain
\be
\langle A \rangle= \frac{N^2}{16} \left[1 -\frac{32 \pi^2}{3 \pt^2} + 
\frac{64}{\pt^2} 
\mathrm{Li}_2\left(1-\alpha \right) \right] + \mathcal{O}(e^{-\pt/8}).
\label{convexasymp}
\ee
When $J=0$, the last term on the right hand side 
of Eq.~(\ref{convexasymp}) is
zero and the relation reduces to 
\be
\langle A \rangle= \frac{N^2}{16} \left[1 -\frac{32 \pi^2}{3 \pt^2} \right] 
+ \mathcal{O}(e^{-\pt/8}), ~~J=0.
\ee

\section{\label{sec4} Column-Convex Polygons }

\begin {figure}
\includegraphics[width=\columnwidth]{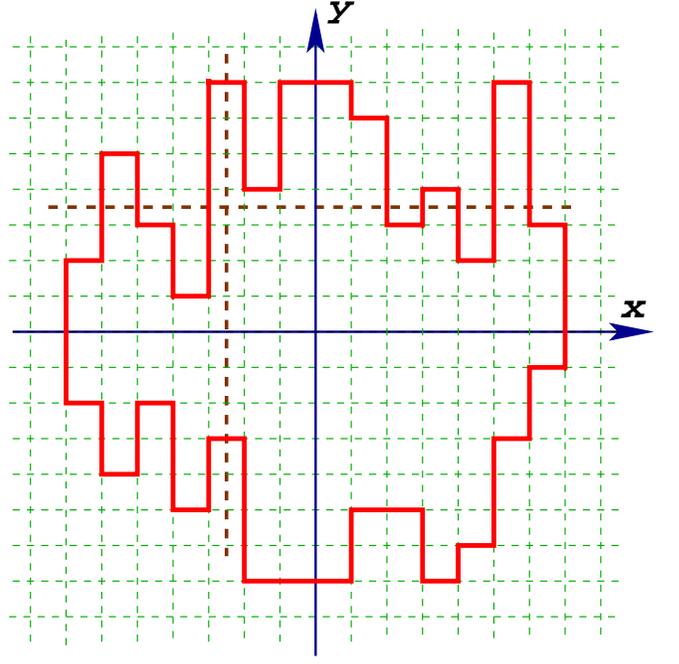}
\caption{\label{columnconvexfig} A schematic diagram of a column-convex 
polygon. Any vertical line intersects the convex polygon at either 
$0$ or $2$ points.
}
\end {figure}

In this section, we calculate the shape and area of a column-convex polygon
when $\pt > 0$. Column-convex polygons 
are those polygons which have exactly $0$ or $2$ intersections with any 
vertical line drawn through the midpoints of the edges of the lattice. 
There, is however, no such restriction in the horizontal direction (see
Fig.~\ref{columnconvexfig}). We
calculate the area by determining the shape of the column-convex polygon that 
minimizes the free energy for a fixed perimeter.

The perimeter of a column-convex polygon has no simple relation to
its bounding box. We thus introduce a chemical potential $\mu$ that couples
to the perimeter $N$. Consider a shape $y(x)$ with endpoints at $(-\beta N,0)$
and $(\beta N,0)$.
The free energy functional for
this curve $y(x)$ is given by
\be
\mathcal{L}[y(x)] = \int_{-\beta N}^{\beta N} dx \, \sigma(y')\sqrt{1+y'^2} 
- \frac{\pt}{N} \int_{-\beta N}^{\beta N} dx~ y,
\label{row_convexL}
\ee
where, $\pt$ is the scaled pressure, $\pt = pN$. As
before, $\sigma(y')$ represents the free energy per unit length associated 
with a slope $y'$. The Euler-Lagrange equation then gives the shape of the
curve equilibrium curve $y(x)$.

The free energy may be calculated as follows. Consider all paths
starting from (0,0) to ($x,y$). 
Let $C(x,y)$ be the weighted sum of all paths. Then, we obtain
\be
C(x,y) = \sum_{y_1,y_2,\ldots,y_X} \prod_i W_i ~\delta\left[\sum y_i - Y\right] ,
\label{row_convex_cxy}
\ee
where $W_i$ is the weight associated with the 
$i^{th}$ step and equals 
\be
W_i = \mu^{|y_i|} \left[ \alpha(1-\delta_{y_i,0})+\delta_{y_i,0} \right] \mu.
\ee

Following the steps outlined previously, we 
convert the $\delta$-function in Eq.~(\ref{row_convex_cxy}) into
an integral, obtaining
\be
C(x,y) =  \frac{1}{2 \pi} \int_0^{2 \pi} ds e^{-isy} \left[ f(\mu,\alpha,s) \mu
\right]^x ,
\label{row_convex_cxy1}
\ee
where
\be
f(\mu,\alpha,s) =
\frac{1+(1-2 \alpha)\mu^2+\mu(\alpha-1)(e^{is}+e^{-is})}
{(1-\mu e^{is})(1-\mu e^{-is})} .
\ee

When $y \gg 1$, Eq.~(\ref{row_convex_cxy1}) may be evaluated by
the saddle point method. Denoting $y/x$ by $y'$, we obtain
\be
\sigma(y') = \frac{1}{\sqrt{1+y'^2}} \left[ is_0 y' - \ln \mu - 
\ln f(\mu,\alpha,s_0)\right] ,
\ee
where $s_0$ is the saddle point and is given by,
\be
\frac{d}{d s_0} \ln f(\mu,\alpha,s_0) = i y' . 
\label{s0}
\ee
Substituting the expression for $\sigma(y')$ into the Euler-Lagrange equation 
(Eq.~(\ref{eq:euler})) and using Eq.~(\ref{s0}) we integrate once to obtain an
equation for $y'$. This gives
\bea
y' &=& \frac{\mu (\alpha-1) (c e^{-\pt x/N} - c^{-1} e^{\pt x/N})}
{1+(1-2\alpha)\mu^2 + \mu(\alpha-1)(c e^{-\pt x/N} + c^{-1} e^{\pt x/N})} 
\nonumber \\
&& + \frac{\mu c e^{-\pt x/N}}{1-\mu c e^{-\pt x/N}}
- \frac{\mu c^{-1} e^{\pt x/N}}{1-\mu c^{-1} e^{\pt x/N}}.
\eea
The constant of integration $c$ is fixed by the condition that the
slope of the equilibrium curve is $0$ ($y' = 0$) at $x=0$. This gives $c=1$. 
Then we can integrate
once more to obtain the equation of the equilibrium curve as
\bea
\label{columnconvexshapeeq}
\lefteqn{Y(X) =}     \\ \nonumber
&& \!\!  -\frac{c_1}{\pt} + \frac{1}{p} \ln \left[ \frac{(1-\mu e^{\pt X}) (1- \mu e^{-\pt X})}
{1+(1-\alpha)\mu^2 +\mu (\alpha-1)(e^{\pt X}+e^{-\pt X})}\right]. 
\eea
As before, $X$ and $Y$ are defined as $X=x/N$ and $Y=y/N$.
The constant of integration $c_1$ is fixed by the boundary condition
$y(x=\beta N)=0$. This gives,
\be
c_1 = \ln \frac{(1-\mu e^{\pt \beta}) (1- \mu e^{-\pt \beta})}
{1+(1-\alpha)\mu^2 +\mu (\alpha-1)(e^{\pt \beta}+e^{-\pt \beta})} .
\label{ccconstant}
\ee

The parameter $\beta$ that determines the endpoint of the
curve is still undetermined. It is chosen to be that $\beta$ that
minimizes the free energy.
The Lagrangian $\mathcal{L}_0$ for this curve $Y(X)$ is  given by
substituting Eqs.~(\ref{columnconvexshapeeq}) and
(\ref{ccconstant})  into Eq.~(\ref{row_convexL}):
\bea
\label{eq:L}
\lefteqn{\mathcal{L}_0 = 2 \beta N \int_0^1 dz} \\ \nonumber
&& \left[ \ln \frac{(1-\mu e^{\pt z \beta}) (1- \mu e^{-\pt z \beta})}
{1+(1-\alpha)\mu^2 +\mu (\alpha-1)(e^{\pt z \beta}+e^{-\pt z \beta})} - 
\ln \mu \right] .
\eea
The parameter 
 $\beta$ satisfies the equation
\be
\frac{d \mathcal{L}_0}{d \beta} = 0 .
\ee
This gives
\bea
\lefteqn{ g(\mu,J) \equiv e^{\beta_0 \pt} = 
\frac{1-\mu+\mu^2-\mu^3 (1-2 \alpha)}{2 \mu [1+\mu (\alpha-1)]} } \\ \nonumber
&&+ \frac{\sqrt{(1-\mu^2) [1-2\mu+\mu^2 (1-2\alpha)] [1-\mu^2 (1-2\alpha)]}}
{2 \mu [1+\mu (\alpha-1)]} .
\eea

\begin {figure}
\includegraphics[width=\columnwidth]{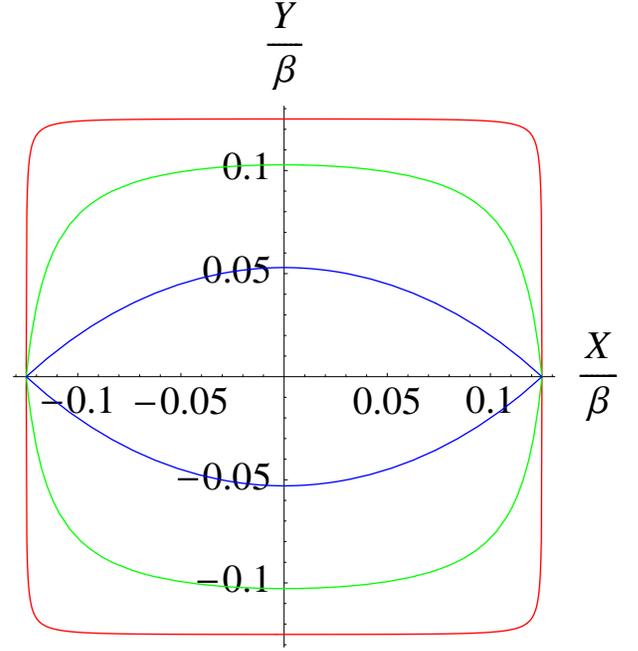}
\caption{\label{columnconvexshape} Shape of the convex polygon as obtained from
Eq.~(\ref{columnconvexshapeeq}) for polygons with $J=1$. The
different shapes correspond to pressure values $\pt=3.0,10.0,50.0$, with
the outer curve corresponding to the largest pressure. Both the
$X$ and $Y$ axes are scaled by $\beta$.}
\end {figure}

The chemical potential $\mu$ is determined by the constraint that
total perimeter is $N$. This is equivalent to 
\be
\mu \frac{d \mathcal{L}_0}{d \mu} = - \frac{N}{2} .
\ee
$\mu$ then satisfies the equation,
\bea
\lefteqn{-\frac{\pt}{4} = \ln \frac{1-\mu g}{g-\mu} -\ln g } \\ \nonumber
&+& \frac{2-a}{\sqrt{a^2-b^2}} 
\ln \frac{(a+b)(g+1)+\sqrt{a^2-b^2}(g-1)}{(a+b)(g+1)-\sqrt{a^2-b^2}(g-1)} ,
\label{eq:mu0}
\eea
where $a$ and $b$ are given by
\bea
a &=& 1+(1-2\alpha)\mu^2, \\
b &=& 2 \mu (\alpha-1) .
\eea
This solves the equilibrium macroscopic shape completely.
The shapes given by Eq.~(\ref{columnconvexshapeeq})
are plotted in Fig.~\ref{columnconvexshape} for 
column-convex polygons with $J=1.0$. 

We now determine the asymptotic behavior of area
when $\pt \gg 1$. This corresponds to the limit $\mu \rightarrow 0$.
In this limit, $g(\mu,J)$ can be expanded as 
\be
g(\mu,J) = \frac{1}{\mu} - \alpha +\alpha(\alpha-1)\mu 
- \alpha (\alpha-1)^2 \mu^2 + \mathcal{O}(\mu^3) .
\label{eq:g}
\ee
and Eq.~(\ref{eq:mu0})
reduces to
\be
\mu = e^{-\pt/8} + \mathcal{O}(e^{-\pt/4}). \label{eq:mu}
\ee
Substituting the values of $g(\mu,J)$ and $\mu$ from Eqs.~(\ref{eq:g})
and (\ref{eq:mu}) into Eq.~(\ref{eq:L}), we can obtain the Lagrangian in the
($\mu,N$) coordinates to be
\be
\mathcal{L}_0(\mu,N) = \frac{\pt N}{32} + 
\frac{2N}{\pt} [\li_2(1-\alpha) -\frac{\pi^2}{6}] + \mathcal{O}(e^{-\pt/8}) .
\ee
The Lagrangian in the ($\pt,N$) coordinates can then be obtained by a Legendre
transformation as
\bea
\mathcal{L}_0(\pt,N) &=& \mathcal{L}_0(\mu,N) + \ln(\mu) \frac{N}{2}, \\
&=& \frac{-\pt N}{32} + 
\frac{2N}{\pt} [\li_2(1-\alpha) -\frac{\pi^2}{6}] + \mathcal{O}(e^{-\pt/8}) ,\nonumber \\
\eea
when $\pt \gg 1$.
The area enclosed by the column-convex polygon is
\bea
A &= &- 2 N \frac{\partial \mathcal{L}_0}{\partial \pt}, \label{area_defn}\\
& =&
\frac{N^2}{16} \left[1- \frac{32 \pi^2}{3 \pt^2} + \frac{64}{\pt^2} \li_2(1-\alpha)\right]
+ \mathcal{O}(e^{-\pt/8}), \nonumber \\
\label{columnconvexasymp}
\eea
where the factor $2$ in Eq.~(\ref{area_defn}) accounts for the lower
half plane. Interestingly, Eq.~(\ref{columnconvexasymp}) is identical 
to Eq.~(\ref{convexasymp}) which is the asymptotic area expression for
convex polygons.

\section{\label{sec5} Self-avoiding and self-intersecting polygons}

In this section, we study the asymptotic behavior of 
self-avoiding and self-intersecting
polygons.  An  analytic calculation along the lines of those
presented for convex and
column-convex polygons is not possible for self-avoiding polygons.
However, we argue as follows:
Convex polygons have no overhangs and the shape has four cusps. Introducing
overhangs in one direction gives column convex polygons, reducing
the number of cusps 
by two. Remarkably, the asymptotic behavior of the
area in the column-convex case [Eq.~(\ref{columnconvexasymp}] 
coincides with that for convex polygons [Eq.~(\ref{convexasymp})]. 
It is therefore plausible that introducing 
overhangs in both directions does not affect the asymptotic 
behavior of the area,
but merely removes the remaining two cusps, yielding a smooth 
shape. We therefore conjecture
that the asymptotic behavior of the area of self-avoiding
polygons is given by 
\be
\langle A \rangle= \frac{N^2}{16} \left[1 -\frac{32 \pi^2}{3 \pt^2} + 
\frac{64}{\pt^2} \mathrm{Li}_2\left(1-\alpha \right) \right] ,
~~~\pt \gg 1 .
\label{sapasymp}
\ee

For self-intersecting polygons in the inflated phase, it is expected 
that the typical shape of the polygon does not intersect itself.
Therefore, we argue that the area of self-intersecting polygons should also have the same
asymptotic behavior as in Eq.~(\ref{sapasymp}).

These conjectures may be verified numerically. When
$J=0$, the area of self-avoiding polygons may be obtained using
exact enumeration data on the square lattice
\cite{jensen03}. This data is available for lengths up
to $N=90$ \cite{jensensap}. When $J\neq 0$, there is no
exact enumeration data available. We therefore resort
to Monte Carlo simulations. The Monte Carlo algorithm
consists of a combination of global reflection and
inversion moves \cite{madras90}.  The system size used
was $N=800$. 

For self-intersecting polygons, the area may be computed using
exact enumeration methods.
We briefly describe the algorithm for the case
$J=0$.  The generalization to non-zero $J$ can be found
in Ref.~\cite{mitra07}. Consider a random walk starting from
the origin and taking steps in one of the four
possible directions. For each step in the positive
(negative) $x$-direction, we assign a weight $e^{-p
y}$ ($e^{py}$), where $y$ is the ordinate of the
walker. The weight is then $e^{pA}$
for a closed walk enclosing an area $A$.
Let $T_N(x,y)$ be the weighted sum of  all $N$-step
walks from $(0,0)$ to $(x,y)$.
It obeys  the recursion relation,
\bea
T_{N+1} (x,y) &=& e^{-p y} T_N (x-1,y) + e^{p y} T_N (x+1,y) \nonumber \\
&&+ T_N (x,y-1) + T_N (x,y+1),
\eea
with the initial condition
\be
T_0 (x,y) = \delta_{x,0} \delta_{y,0}.
\ee
Finally, $T_N(0,0)$ gives the partition function of the self-intersecting
polygons on a
lattice.
We used exact enumeration data
up to $N=150$. 

\begin {figure}
\includegraphics[width=\columnwidth]{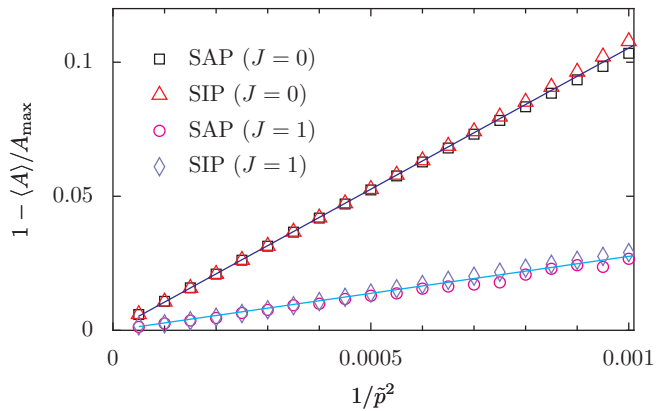}
\caption{\label{sipdata} 
The asymptotic behavior of the area for self-avoiding (SAP) and
self-intersecting polygons (SIP).
The solid lines correspond to the theoretical prediction of 
Eq.~(\ref{sapasymp}). The data is
in good agreement with
Eq.~(\ref{sapasymp}).}
\end {figure}

In the case of all exact enumeration
data, for each pressure point, we extrapolated to
$N\rightarrow \infty$ using finite size scaling.
The results of the numerical
analysis is shown in Fig.~\ref{sipdata}. The numerical 
data agree very well with the theoretical prediction.

\section{\label{sec6} Conclusions}

We now summarize the basic results of this paper. We 
studied the asymptotic behavior of area for large pressures
for a class of polygons. For convex and column-convex polygons, we 
calculated the area using the Wulff construction. The asymptotic
behavior was observed to coincide for both classes of polygons. We therefore
conjectured that overhangs are not important in the inflated regime, and hence
that self avoiding polygons should have the same asymptotic behavior. This
was verified numerically. We also showed numerically that self intersections
were irrelevant in the inflated regime. These results continue to
remain valid in the presence of a non-zero bending rigidity.

Interestingly, the asymptotic behaviour for continuum ring polymers
differs from that of lattice polygons. 
In the continuum, the analogous relation for
the area of pressurized rings is asymptotically \cite{haleva06,mitra07}
\be
\frac{\langle A \rangle}{A_{max}} \longrightarrow 1 - \frac{2 \pi}{\pt} ,
~~~ \pt \gg 1 .
\ee
This difference between continuum and lattice models
is physically sensible in the expanded limit, since curvature
in the lattice case is concentrated in local regions with
$\pi/2$ bends but is delocalized along the contour in the
continuum case.

While our conjectured result for self-avoiding polygons is in good
agreement with numerical data, it would be of interest to have a
mathematically rigorous derivation of this result.
It may be possible to extend the methods of 
Ref.~\cite{prellberg99} to this problem.


\begin{thebibliography}{26}
\expandafter\ifx\csname natexlab\endcsname\relax\def\natexlab#1{#1}\fi
\expandafter\ifx\csname bibnamefont\endcsname\relax
  \def\bibnamefont#1{#1}\fi
\expandafter\ifx\csname bibfnamefont\endcsname\relax
  \def\bibfnamefont#1{#1}\fi
\expandafter\ifx\csname citenamefont\endcsname\relax
  \def\citenamefont#1{#1}\fi
\expandafter\ifx\csname url\endcsname\relax
  \def\url#1{\texttt{#1}}\fi
\expandafter\ifx\csname urlprefix\endcsname\relax\def\urlprefix{URL }\fi
\providecommand{\bibinfo}[2]{#2}
\providecommand{\eprint}[2][]{\url{#2}}

\bibitem[{\citenamefont{Leibler et~al.}(1987)\citenamefont{Leibler, Singh, and
  Fisher}}]{leibler87}
\bibinfo{author}{\bibfnamefont{S.}~\bibnamefont{Leibler}},
  \bibinfo{author}{\bibfnamefont{R.~R.~P.} \bibnamefont{Singh}},
  \bibnamefont{and} \bibinfo{author}{\bibfnamefont{M.~E.}
  \bibnamefont{Fisher}}, \bibinfo{journal}{Phys. Rev. Lett.}
  \textbf{\bibinfo{volume}{59}}, \bibinfo{pages}{1989} (\bibinfo{year}{1987}).

\bibitem[{\citenamefont{Fisher et~al.}(1991)\citenamefont{Fisher, Guttmann, and
  Whittington}}]{fisher91}
\bibinfo{author}{\bibfnamefont{M.~E.} \bibnamefont{Fisher}},
  \bibinfo{author}{\bibfnamefont{A.~J.} \bibnamefont{Guttmann}},
  \bibnamefont{and} \bibinfo{author}{\bibfnamefont{S.~G.}
  \bibnamefont{Whittington}}, \bibinfo{journal}{J. Phys. A}
  \textbf{\bibinfo{volume}{24}}, \bibinfo{pages}{3095} (\bibinfo{year}{1991}).

\bibitem[{\citenamefont{Satyanarayana and
  Baumgaertner}(2004)}]{satyanarayana04}
\bibinfo{author}{\bibfnamefont{S.~V.~M.} \bibnamefont{Satyanarayana}}
  \bibnamefont{and}
  \bibinfo{author}{\bibfnamefont{A.}~\bibnamefont{Baumgaertner}},
  \bibinfo{journal}{J. Chem. Phys.} \textbf{\bibinfo{volume}{121}},
  \bibinfo{pages}{4255} (\bibinfo{year}{2004}).

\bibitem[{\citenamefont{van Faassen}(1998)}]{faassen98}
\bibinfo{author}{\bibfnamefont{E.}~\bibnamefont{van Faassen}},
  \bibinfo{journal}{Physica A} \textbf{\bibinfo{volume}{255}},
  \bibinfo{pages}{251} (\bibinfo{year}{1998}).

\bibitem[{\citenamefont{Privman and Svrakic}(1989)}]{privman}
\bibinfo{author}{\bibfnamefont{V.}~\bibnamefont{Privman}} \bibnamefont{and}
  \bibinfo{author}{\bibfnamefont{N.}~\bibnamefont{Svrakic}},
  \emph{\bibinfo{title}{Directed Models of Polymers, Interfaces, and Clusters:
  Scaling and Finite-Size Properties}} (\bibinfo{publisher}{Springer-Verlag},
  \bibinfo{year}{1989}).

\bibitem[{\citenamefont{Rajesh and Dhar}(2005)}]{rajesh05}
\bibinfo{author}{\bibfnamefont{R.}~\bibnamefont{Rajesh}} \bibnamefont{and}
  \bibinfo{author}{\bibfnamefont{D.}~\bibnamefont{Dhar}},
  \bibinfo{journal}{Phys. Rev. E} \textbf{\bibinfo{volume}{71}},
  \bibinfo{pages}{016130} (\bibinfo{year}{2005}).

\bibitem[{\citenamefont{Bousquet-Melou}(1996)}]{bousquet96}
\bibinfo{author}{\bibfnamefont{M.}~\bibnamefont{Bousquet-Melou}},
  \bibinfo{journal}{Discrete Math.} \textbf{\bibinfo{volume}{154}},
  \bibinfo{pages}{1} (\bibinfo{year}{1996}).

\bibitem[{\citenamefont{van Rensburg}(2000)}]{rensburgbook}
\bibinfo{author}{\bibfnamefont{E.~J.~J.} \bibnamefont{van Rensburg}},
  \emph{\bibinfo{title}{The Statistical Mechanics of Interacting Walks,
  Polygons, Animals and Vesicles}} (\bibinfo{publisher}{Oxford University
  Press}, \bibinfo{year}{2000}).

\bibitem[{\citenamefont{Lin}(1991)}]{lin91}
\bibinfo{author}{\bibfnamefont{K.~Y.} \bibnamefont{Lin}}, \bibinfo{journal}{J.
  Phys. A.} \textbf{\bibinfo{volume}{24}}, \bibinfo{pages}{2411}
  (\bibinfo{year}{1991}).

\bibitem[{\citenamefont{Bousquet-Melou}(1992{\natexlab{a}})}]{bousquet92_1}
\bibinfo{author}{\bibfnamefont{M.}~\bibnamefont{Bousquet-Melou}},
  \bibinfo{journal}{J. Phys. A.} \textbf{\bibinfo{volume}{25}},
  \bibinfo{pages}{1925} (\bibinfo{year}{1992}{\natexlab{a}}).

\bibitem[{\citenamefont{Bousquet-Melou}(1992{\natexlab{b}})}]{bousquet92_2}
\bibinfo{author}{\bibfnamefont{M.}~\bibnamefont{Bousquet-Melou}},
  \bibinfo{journal}{J. Phys. A.} \textbf{\bibinfo{volume}{25}},
  \bibinfo{pages}{1935} (\bibinfo{year}{1992}{\natexlab{b}}).

\bibitem[{\citenamefont{Brak and Guttmann}(1990)}]{brak90_1}
\bibinfo{author}{\bibfnamefont{R.}~\bibnamefont{Brak}} \bibnamefont{and}
  \bibinfo{author}{\bibfnamefont{A.~J.} \bibnamefont{Guttmann}},
  \bibinfo{journal}{J. Phys. A} \textbf{\bibinfo{volume}{23}},
  \bibinfo{pages}{4581} (\bibinfo{year}{1990}).

\bibitem[{\citenamefont{Jensen}(2003)}]{jensen03}
\bibinfo{author}{\bibfnamefont{I.}~\bibnamefont{Jensen}}, \bibinfo{journal}{J.
  Phys. A} \textbf{\bibinfo{volume}{36}}, \bibinfo{pages}{5731}
  (\bibinfo{year}{2003}).

\bibitem[{\citenamefont{Richard et~al.}(2001)\citenamefont{Richard, Guttmann,
  and Jensen}}]{richard01}
\bibinfo{author}{\bibfnamefont{C.}~\bibnamefont{Richard}},
  \bibinfo{author}{\bibfnamefont{A.~J.} \bibnamefont{Guttmann}},
  \bibnamefont{and} \bibinfo{author}{\bibfnamefont{I.}~\bibnamefont{Jensen}},
  \bibinfo{journal}{J. Phys. A} \textbf{\bibinfo{volume}{34}},
  \bibinfo{pages}{L495} (\bibinfo{year}{2001}).

\bibitem[{\citenamefont{Cardy}(2001)}]{cardy01}
\bibinfo{author}{\bibfnamefont{J.}~\bibnamefont{Cardy}}, \bibinfo{journal}{J.
  Phys. A} \textbf{\bibinfo{volume}{34}}, \bibinfo{pages}{L665}
  (\bibinfo{year}{2001}).

\bibitem[{\citenamefont{Richard}(2002)}]{richard02}
\bibinfo{author}{\bibfnamefont{C.}~\bibnamefont{Richard}}, \bibinfo{journal}{J.
  Stat. Phys.} \textbf{\bibinfo{volume}{108}}, \bibinfo{pages}{459}
  (\bibinfo{year}{2002}).

\bibitem[{\citenamefont{Rudnick and Gaspari}(1991)}]{rudnick91}
\bibinfo{author}{\bibfnamefont{J.}~\bibnamefont{Rudnick}} \bibnamefont{and}
  \bibinfo{author}{\bibfnamefont{G.}~\bibnamefont{Gaspari}},
  \bibinfo{journal}{Science} \textbf{\bibinfo{volume}{252}},
  \bibinfo{pages}{422} (\bibinfo{year}{1991}).

\bibitem[{\citenamefont{Gaspari et~al.}(1993)\citenamefont{Gaspari, Rudnick,
  and Beldjenna}}]{gaspari93}
\bibinfo{author}{\bibfnamefont{G.}~\bibnamefont{Gaspari}},
  \bibinfo{author}{\bibfnamefont{J.}~\bibnamefont{Rudnick}}, \bibnamefont{and}
  \bibinfo{author}{\bibfnamefont{A.}~\bibnamefont{Beldjenna}},
  \bibinfo{journal}{J. Phys. A} \textbf{\bibinfo{volume}{26}},
  \bibinfo{pages}{1} (\bibinfo{year}{1993}).

\bibitem[{\citenamefont{Haleva and Diamant}(2006)}]{haleva06}
\bibinfo{author}{\bibfnamefont{E.}~\bibnamefont{Haleva}} \bibnamefont{and}
  \bibinfo{author}{\bibfnamefont{H.}~\bibnamefont{Diamant}},
  \bibinfo{journal}{Eur. Phys. J. E} \textbf{\bibinfo{volume}{19}},
  \bibinfo{pages}{461} (\bibinfo{year}{2006}).

\bibitem[{\citenamefont{Mitra et~al.}()\citenamefont{Mitra, Menon, and
  Rajesh}}]{mitra07}
\bibinfo{author}{\bibfnamefont{M.~K.} \bibnamefont{Mitra}},
  \bibinfo{author}{\bibfnamefont{G.~I.} \bibnamefont{Menon}}, \bibnamefont{and}
  \bibinfo{author}{\bibfnamefont{R.}~\bibnamefont{Rajesh}},
  \bibinfo{howpublished}{preprint arXiv:0708.3318}.

\bibitem[{\citenamefont{Prellberg and Owczarek}(1999)}]{prellberg99}
\bibinfo{author}{\bibfnamefont{T.}~\bibnamefont{Prellberg}} \bibnamefont{and}
  \bibinfo{author}{\bibfnamefont{A.~L.} \bibnamefont{Owczarek}},
  \bibinfo{journal}{Commun. Math. Phys.} \textbf{\bibinfo{volume}{201}},
  \bibinfo{pages}{493} (\bibinfo{year}{1999}).

\bibitem[{\citenamefont{Maggs et~al.}(1990)\citenamefont{Maggs, Leibler,
  Fisher, and Camacho}}]{maggs90}
\bibinfo{author}{\bibfnamefont{A.~C.} \bibnamefont{Maggs}},
  \bibinfo{author}{\bibfnamefont{S.}~\bibnamefont{Leibler}},
  \bibinfo{author}{\bibfnamefont{M.~E.} \bibnamefont{Fisher}},
  \bibnamefont{and} \bibinfo{author}{\bibfnamefont{C.~J.}
  \bibnamefont{Camacho}}, \bibinfo{journal}{Phys. Rev. A}
  \textbf{\bibinfo{volume}{42}}, \bibinfo{pages}{691} (\bibinfo{year}{1990}).

\bibitem[{\citenamefont{Fisher}(1966)}]{fisher66}
\bibinfo{author}{\bibfnamefont{M.~E.} \bibnamefont{Fisher}},
  \bibinfo{journal}{J. Chem. Phys.} \textbf{\bibinfo{volume}{44}},
  \bibinfo{pages}{616} (\bibinfo{year}{1966}).

\bibitem[{\citenamefont{Jensen}()}]{jensensap}
\bibinfo{author}{\bibfnamefont{I.}~\bibnamefont{Jensen}},
  \emph{\bibinfo{title}{Number of sap of given perimeter and any area}},
  \urlprefix\url{http://www.ms.unimelb.edu.au/~iwan/polygons/series/sqsap_peri%
m_area.ser}.

\bibitem[{\citenamefont{Rottman and Wortis}(1984)}]{rottmannwortis84}
\bibinfo{author}{\bibfnamefont{C.}~\bibnamefont{Rottman}} \bibnamefont{and}
  \bibinfo{author}{\bibfnamefont{M.}~\bibnamefont{Wortis}},
  \bibinfo{journal}{Phys. Rep.} \textbf{\bibinfo{volume}{103}},
  \bibinfo{pages}{59} (\bibinfo{year}{1984}).

\bibitem[{\citenamefont{Madras et~al.}(1990)\citenamefont{Madras, Orlitsky, and
  Shepp}}]{madras90}
\bibinfo{author}{\bibfnamefont{N.}~\bibnamefont{Madras}},
  \bibinfo{author}{\bibfnamefont{A.}~\bibnamefont{Orlitsky}}, \bibnamefont{and}
  \bibinfo{author}{\bibfnamefont{L.~A.} \bibnamefont{Shepp}},
  \bibinfo{journal}{J. Stat. Phys.} \textbf{\bibinfo{volume}{58}},
  \bibinfo{pages}{159} (\bibinfo{year}{1990}).

\end{thebibliography}
\end{document}